\documentclass[useAMS,usenatbib]{mn2e}
\usepackage{graphics,graphicx}
\usepackage{amssymb}
\usepackage{latexsym}
\usepackage{color}
\usepackage{epstopdf}
\usepackage{marvosym}
\usepackage{multirow}

\newcommand{\jcp}{J. Chem. Phys.}
\newcommand{\aap}{A\&A}
\newcommand{\apj}{ApJ}

\newcommand{\araa}{ARA\&A}
\newcommand{\mnras}{MNRAS}

\begin{document}

\title[Electron-impact excitation of diatomic hydride
  cations]{Electron-impact excitation of diatomic hydride cations 
  II: OH$^+$ and SH$^+$} 
\author[James R Hamilton, Alexandre Faure
  and Jonathan Tennyson]{James R Hamilton$^1$\thanks{E-mail:
    james.hamilton@ucl.ac.uk}, Alexandre Faure$^2$\thanks{E-mail:
    alexandre.faure@univ-grenoble-alpes.fr} and Jonathan
  Tennyson$^1$\thanks{E-mail: j.tennyson@ucl.ac.uk}\\ 
$^1$ Department
  of Physics and Astronomy, University College, London, Gower St.,
  London WC1E 6BT, UK\\ 
$^2$ Univ. Grenoble Alpes, CNRS, IPAG, F-38000
  Grenoble, France}

\date{Accepted ? Received ?}

\maketitle

\begin{abstract}

{\bf R}-matrix calculations combined with the
adiabatic-nuclei-rotation and Coulomb-Born approximations are used to
compute electron-impact rotational rate coefficients for two
open-shell diatomic cations of astrophysical interest: the hydoxyl and
sulphanyl ions, OH$^+$ and SH$^+$. Hyperfine resolved rate
coefficients are deduced using the infinite-order-sudden
approximation. The propensity rule $\Delta F=\Delta j=\Delta N=\pm 1$
is observed, as is expected for cations with a large dipole moment. A
model for OH$^+$ excitation in the Orion Bar photon-dominated region
(PDR) is presented which nicely reproduces {\it Herschel} observations
for an electron fraction $x_e=10^{-4}$ and an OH$^+$ column density of
$3\times 10^{13}$~cm$^{-2}$. Electron impact electronic excitation
cross sections and rate coefficients for the ions are also presented.

\end{abstract}


\section{Introduction}

Cross sections for electron collisions with molecular ions can be very
large ($>$1000~\AA$^2$). If the ion in question contains a permanent dipole
moment, the electron-impact rotational excitation rate coefficients
far exceed those of H and H$_2$ meaning that in comparatively
electron-rich regions, electron collisions can become the dominant
excitation process. Rotational rate coefficients have already been
used to quantify interstellar electron densities \citep{jt392,jt565,jt617}, 
but the
rate coefficients for many key species remain unknown. In this paper
we consider (de)excitation of the hydoxyl and sulphanyl ions: OH$^+$
and SH$^+$, respectively. The species both have electronic ground
states of $^3\Sigma^-$ symmetry which adds an extra complication as
the rotational levels display fine structure due to the electron spin
of the two unpaired electrons and hyperfine structure due to the
nuclear spin of the hydrogen atom.

Both OH$^+$ and SH$^+$ were only detected in the interstellar medium
within the last decade; OH$^+$ being first observed by
\citet{Wyrowski2010} and SH$^+$ by \citet{benz2010} and
\citet{Menten2011}. However, the ions are now known to be widespread
\citep{gerin2016}. In particular OH$^+$ has now been found in a
variety of locations including translucent interstellar clouds
\citep{kre2010,gupta10} and both OH$^+$ and SH$^+$ have been recently observed
in absorption across the $z=0.89$ molecular absorber towards PKS
1830-211 \citep{muller2016,muller2017}. They have been also detected
in emission in dense photon-dominated regions where electron collision
processes are thought to be important \citep{tak2013,nagy13}. A
number of these observations resolve the fine (and sometimes
hyperfine) structure in the transitions
\citep{benz2010,gerin2010,godard2012,nagy13}.

To date there is only one laboratory measurement of electron-impact
rotational rate coefficients for a molecular ion was by \citet{jt465} for HD$^+$;
this experiment actually measured de-excitation and only gave enough
information to show agreement with the theoretical predictions. This
means that thus far astronomically important electron-impact
rotational rate coefficients for molecular ions have all been computed
\citep{jt271,jt313}. In a recent paper \citep{jt617}, we used improved
theory to compute rotational rate coefficients for three closed shell
hydride cations, ArH$^+$, CH$^+$ and HeH$^+$; these hydrides were
chosen due to their significant role in the interstellar medium (ISM),
see \citet{jt688} for example. In this work, electron-impact rate
coefficients are calculated for the open-shell ions OH$^+$ and
SH$^+$. \textbf{R}-matrix calculations are combined with the
adiabatic-nuclei-rotation (ANR) approximation to produce rotational cross
sections at electron energies below 5~eV. We also present electron
impact electronic excitation cross sections for the two ions
considered.  While these are unlikely to be important for models of
interstellar medium, OH$^+$ can be found in planetary ionospheres
\citep{fox15}, and cometary coma \citep{Nordholt03,Haider05,rubin09},
as well as around Enceladus \citep{gupta10}. In these environments
electron impact electronic excitation may well be important.

Section~2 describes the {\bf
  R}-matrix calculations  and the procedure used to
derive the cross-sections and rate coefficients is briefly
introduced. In Section 3, we present and discuss the calculated rate
coefficients. A model for the excitation of OH$^+$ in the Orion bar
photon-dominated region (PDR) is also presented in
Section~4. Conclusions are summarized in Section~5.

\section{R-matrix calculations}

Inelastic electron collision calculations with molecular ions OH$^+$ and SH$^+$ were
performed using the R-matrix method
\citep{jt474} within the Quantemol-N \citep{jt416} expert system to
run the UK molecular {\bf R}-matrix codes (\texttt{UKRMol})
\citep{jt518}. Details follow closely the calculations performed in our previous paper
\citep{jt617}, denoted I below, and are not repeated here.
The calculations produce \mbox{{\bf T}-matrices} which are processed by
electron-impact rotational excitation code \texttt{ROTIONS}
\citep{jt226}, which employs the Coulomb-Born approximation to include
the effects of high partial waves \citep{np82}.  In particular,
$\Delta N = 1$ transitions ($N$ is the molecular ion rotational angular momentum)
are strongly influenced by the long-range dipole moment
and \texttt{ROTIONS} uses the Coulomb-Born approximation to include
the contributions of partial waves with $\ell > 4$. These long-range effects are
unimportant for transitions with $\Delta N>1$ \citep{jt271}. Experimental
 values of the dipole moments were used
in these calculations where available. 

\subsection{OH$^+$}

The OH$^+$ target was represented using an augmented aug-cc-pVTZ GTO
basis set. The use of augmented basis sets improves the treatment of
the more diffuse orbitals for the excited states in the calculation.
The ground state of OH$^+$ is X~$^3\Sigma^-$ which has the configuration
[1~$\sigma$ 2~$\sigma$ 3~$\sigma$]$^6$ [1~$\pi$]$^2$.  The target was
represented using CAS-CI treatment freezing the lowest energy
1~$\sigma^2$ orbital and placing the highest 6 electrons in orbitals
[2-8~$\sigma$, 1-3~$\pi$]$^6$. This target was constructed in an
\textbf{R}-matrix sphere of radius 13~a$_0$. Nine electronically excited
states were used in the close-coupling expansion.

The vertical excitation energies (VEEs) of the excited states of
OH$^+$ calculated using this model at an equilibrium bondlength of 1.0289~\AA\  are given in
Table~\ref{tab:OH+ExStat}, where the VEEs are compared to
 measured values. The VEEs calculated in this work compare
well to the measured adiabatic excitation energies (AEEs). VEEs
naturally exceed AEEs and in this particular case the  A~$^3\Pi$
has a much larger equilibrium bondlength (1.134~\AA) than the
b~$^1\Sigma^+$ state (1.032~\AA), which results in a different order
of the states at $R=1.029$~\AA.
The excited states a~$^1\Delta$,
b~$^1\Sigma^+$ and A~$^3\Pi$ are within the electron energy range of
interest in this investigation. Calculated equilibrium geometry dipole
moment and rotational constant of OH$^+$ are compared to the best
available values in Table~\ref{tab:OH+DipRotData}.

\begin{table}
\caption [Vertical excitation energies for the lowest 5 excited states
  of OH$^+$]{ Vertical excitation energies for the lowest 5 excited
  states of OH$^+$ compared with measured adiabatic excitation
  energies.}
\begin{center}
    \begin{tabular}{lcl}
    \hline\hline
     \multicolumn{1}{ c }{State} & This Work (eV) & Previous (eV)\\ 
    \hline
     X~$^3\Sigma^-$  &  ~0.000 &  0.000 \\ 
     a~$^1\Delta$    &  ~2.509 &  2.190 $^a$\\ 
     b~$^1\Sigma^+$  &  ~3.709 &  3.602 $^b$\\
     A~$^3\Pi$       &  ~3.903 &  3.526 $^b$\\ 
     1~$^1\Pi$       &  ~6.183 &   \\ 
     2~$^1\Sigma^+$  &  11.510 &   \\ 
    \hline
    \hline
    \end{tabular}
\mbox{}\\
{\flushleft
$^a$ \citet{Katsumata1977}\\
$^b$ \citet{Huber1979}\\
}
\end{center}
\label{tab:OH+ExStat}
\end{table}

Isotopic substitution shifts the centre-of-mass and hence, for ionic
system, alters the permanent dipole moment. Oxygen exists in three
isotopes giving $^{16}$OH$^+$, $^{17}$OH$^+$ and $^{18}$OH$^+$.  While
$^{16}$O is the most abundant isotope, the abundance of $^{18}$O is
not negligible with an isotopic ratio $^{16}$O/$^{18}$O=498.7$\pm$0.1
for the Solar System (Vienna Standard Mean Ocean Water value)
\citep{asplund09,Meija2016}. The abundance of $^{17}$O is much lower
with an isotopic ratio $^{16}$O/$^{17}$O=2632$\pm$7
\citep{asplund09,Meija2016}. To our knowledge, only the main
isotopologue $^{16}$OH$^+$ has been detected in the interstellar
medium so far. For this reason the discussion and results presented in
the main paper will be concerned with only $^{16}$OH$^+$ (henceforth
referred to as OH$^+$) but data for the other isotopologues are also
included in the supplementary data to this article.

\begin{table}
\caption [Dipole moment and rotational constant 
calculated for
  OH$^+$]{Dipole moment, $\mu$, and rotational constant, $B$, calculated for 
isotopologues of OH$^+$
  are compared with published values.}
\begin{center}
    \begin{tabular}{cp{0.76cm}p{0.76cm}p{0.76cm}cl}
    \hline\hline
     Property & \multicolumn{3}{ c }{This Work} & & Previous \\ \cline{2-4} \cline{6-6}
     \rule{0pt}{3ex}     & $^{16}$OH$^+$ & $^{17}$OH$^+$ & $^{18}$OH$^+$ & & $^{16}$OH$^+$ \\ \hline
      $\mu$ (D)          &  ~2.252  & ~2.269 & ~2.283                    & & ~2.256 $^a$  \\ 
      B     (cm$^{-1}$)  &  16.796  & 16.737 & 16.685                    & & 16.423 $^b$,  16.422 $^c$   \\
    \hline\hline
    \end{tabular}
\end{center}
\label{tab:OH+DipRotData}
\mbox{}\\
{\flushleft
$^a$ Theory \citet{Werner1983}\\
$^b$ Rotational spectroscopy \citet{Bekooy1985}\\
$^c$ Ultraviolet spectroscopy \citet{Merer1975}\\
}
\end{table}

\subsection{SH$^+$}

The SH$^+$ target was represented using a non augmented Dunning
\mbox{cc-pVTZ} GTO basis set. Unlike OH$^+$, an augmented basis set
could not be used as it gave linear dependence problems and did not
produce smooth results.  The ground state of SH$^+$ has the
configuration \mbox{[1$\sigma$ 2$\sigma$ 3$\sigma$ 1$\pi$ 4$\sigma$
    5$\sigma$]$^{14}$ [2$\pi$]$^2$}.  The target was represented using
CAS-CI treatment freezing electrons of the lowest energy 1-3~$\sigma$
and 1$\pi$ orbitals and placing the highest 6 electrons in orbitals
\mbox{[4-8~$\sigma$, 2-4~$\pi$, 1~$\delta$]$^6$}.  This target was
constructed in an \textbf{R}-matrix sphere of radius 10~a$_0$.  The
VEEs of the excited states of SH$^+$ calculated from this model at the equilibrium bondlength of 1.3744~\AA\ are
given in Table~\ref{tab:SH+ExStat} and compared to published 
values. The VEEs calculated in this work compare well to the measured
VEEs. The calculated equilibrium geometry dipole moment and rotational
constant of SH$^+$ are compared to the best available values in
Table~\ref{tab:SH+DipRotData}.

\begin{table}
  \caption [Vertical excitation energies for the lowest 7 excited
    states of SH$^+$]{ Vertical excitation energies for the lowest 7
    excited states of SH$^+$ compared with measured values.}
\begin{center}
    \begin{tabular}{lcl}
    \hline\hline
  \multicolumn{1}{c}{State} & This Work (eV) & Previous (eV) \\ \hline
          ~~~X~$^3\Sigma^-$ & ~0.000        &               \\ 
          ~~~a~$^1\Delta$   & ~1.472        & ~1.280 $^{a\dag}$,  ~1.340 $^{b\dag}$ \\
          ~~~b~$^1\Sigma^+$ & ~2.517        & ~2.390 $^{a\dag}$,  ~2.390 $^{b\dag}$\\ 
          ~~~A~$^3\Pi$      & ~3.856        & ~3.740 $^{a\dag}$,  ~3.762 $^{c\ddag}$,  ~3.980 $^{b\dag}$ \\
                            &               & ~3.840 $^{b\ddag}$, ~3.709 $^{d\ddag}$ \\ 
          ~~~c~$^1\Pi$      & ~5.200        & ~5.320 $^{a\dag}$,  ~4.722 $^{c\ddag}$, ~5.220 $^{b\dag}$ \\
                            &               & ~4.800 $^{b\ddag}$ \\
          ~~~~~$^5\Sigma^-$ & ~9.817        & ~9.090 $^{a\dag}$ \\ 
          ~~~2~$^3\Pi$      & 10.304        & \\ 
          ~~~2~$^1\Sigma^+$ & 10.373        & 10.489 $^{b\dag}$ \\ 
    \hline\hline
    \end{tabular}
\end{center}
\label{tab:SH+ExStat}
\mbox{}\\
{\flushleft
$^\dag$ Adiabatic value \\
$^\ddag$  Vertical value \\
$^a$ Observed, \citet{Dunlavey1979}\\
$^b$ Calculated, \citet{Bruna1983}\\
$^c$ Observed, \citet{Rostas1984}\\
$^d$ Observed, \citet{Horani1967}\\
}
\end{table}

Sulphur exists as four isotopes giving $^{32}$SH$^+$, $^{33}$SH$^+$,
$^{34}$SH$^+$ and $^{36}$SH$^+$.  While $^{32}$S is the most abundant
isotope, the abundance of $^{34}$S is significant with an isotopic
ratio $^{32}$S/$^{34}$S$\sim$22 for the Solar System
\citep{asplund09,Meija2016}. The abundances of the other isotopes are
much lower with isotopic ratios $^{32}$S/$^{33}$S$\sim$125 and
$^{32}$S/$^{36}$S$\gtrsim$5000 \citep{asplund09,Meija2016}.
Both isotopologues $^{32}$SH$^+$ and $^{34}$SH$^+$ have
been detected in the (extragalactic for $^{34}$SH$^+$) interstellar
medium \citep{muller2017}. The discussion and results presented in the
main paper will be concerned with only $^{32}$SH$^+$ (henceforth
referred to as SH$^+$) but data for the other isotopologues are also
included in the supplementary data to this article.

\begin{table}
\caption [Dipole moment and rotational constant calculated for
  SH$^+$]{Dipole moment and rotational constant calculated for SH$^+$
  compared with published values.}
\begin{center}
    \begin{tabular}{cp{0.76cm}p{0.76cm}p{0.76cm}p{0.76cm}cl}
    \hline\hline
     Property        & \multicolumn{4}{ c }{This Work}                               & & Previous   \\ \cline{2-5} \cline{7-7}
     \rule{0pt}{3ex} & $^{32}$SH$^+$ & $^{33}$SH$^+$ & $^{34}$SH$^+$ & $^{36}$SH$^+$ & & $^{32}$SH$^+$ \\ \hline
     $\mu$ (D)       & 1.388         & 1.394         & 1.394         & 1.394         & & 1.285 $^a$ \\ 
     B (cm$^{-1}$)   & 9.135         & 9.125         & 9.118         & 9.103         & & 9.133 $^b$ \\
    \hline\hline
    \end{tabular}
\end{center}
\label{tab:SH+DipRotData}
\mbox{}\\
{\flushleft
$^a$ Theory, \citet{Senekowitsch1985}\\
$^b$ Empirical, \citet{Muller2014} \\
}
\end{table}

\subsection{Cross-sections and rate coefficients}

Working in C$_{\rm 2v}$ symmetry, each of the above calculations
produces eight fixed-nuclei {\bf T}-matrices for each molecule: the
four symmetries $A_1, A_2, B_1, B_2$ for both doublet and quartet
states of the $N+1$ electron systems. These {\bf T}-matrices are used
to calculate the electronic excitation cross sections using standard
equations \citep{jt474} and, once converted to the C$_{\infty v}$
point group, rotational excitation cross sections using the program
\texttt{ROTIONS} \citep{jt226} using the rotational constants and 
isotope specific dipole moments  given in Tables~\ref{tab:OH+DipRotData} 
and~\ref{tab:SH+DipRotData}. \texttt{ROTIONS} computes the
rotational excitation cross sections for each doublet and quartet
state independently. The total rotational cross sections are thus
obtained as the (weighted) sum of the doublet and quartet cross
sections.

\subsubsection{Electronic transitions}

Electronic excitation cross sections were computed for
collision energies $E_{coll}$ in the range 0.01-5~eV.
We consider electronic transitions from the ground state of each 
cation to all states with electronic thresholds below the 5~eV 
upper limit. The electronic thresholds are calculated 
using the fixed nuclei approximation. Assuming that the
electron velocity distribution is Maxwellian, rate coefficients for
excitation transitions were obtained for temperatures in the range 1
-- 5000~K.

\subsubsection{Rotational transitions}


As in I, we use a combination of  the adiabatic nuclear rotation
(ANR) method  \citep{Chang1970} with  Coulomb-Born completion (for dipolar
transitions only). To allow for threshold effect we used an empirical correction:
below the excitation threshold cross sections are set
to zero, see \citet{jt396} for details. The validity of this approach was confirmed recently for HeH$^+$ where
the full-rovibrational multichannel quantum defect theory (MQDT)
calculations by \cite{curik17} were found in good agreement with the
ANR/Coulomb-Born calculations of I.

Rotational transitions between levels with $N\leq
11$ were considered.  However, transitions were restricted to $\Delta N\leq 8$ owing to
the finite number of partial waves in the {\bf T}-matrices ($\ell
\leq 4$). Rotational excitation cross sections were computed for
collision energies $E_{coll}$ in the range 0.01-5~eV. For 
transitions with a rotational threshold below 0.01 eV, cross sections
were extrapolated down to the threshold using a $1/E_{coll}$
(Wigner's) law, as recommended by \citet{jt396}. Rate coefficients for
excitation transitions were obtained for temperatures in the range 1
-- 3000~K assuming a thermal electron energy distribution.
The principle of detailed balance was used to compute de-excitation rate coefficients.

\subsubsection{Hyperfine transitions}

As discussed in the introduction, the fine and hyperfine structures of
the OH$^+$ and SH$^+$ ions are resolved in astronomical
observations. It is therefore necessary to provide hyperfine-resolved
rate coefficients for these two ions. In the Hund's case (b) coupling
scheme, the fine structure levels are labelled by $(N, j)$ where ${\bf
  j=N+S}$ is the total angular momentum quantum number and $S=1$ is
the electronic spin. The hyperfine structure levels are labelled by
$(N, j, F)$ where ${\bf F=j+I}$ is the hyperfine quantum number and
$I=1/2$ is the nuclear spin of the hydrogen atom. Each rotational
level is thus split into 3 fine-structure levels ($j=N-1, j=N, j=N+1$)
(except $N=0$) and each fine-structure level is in turn split into 2
hyperfine levels ($F=j\pm 1/2$) (except $(N, j)=(1, 0)$). The fine and
hyperfine splittings are $\sim$ 1~cm$^{-1}$ and $\sim$
0.001~cm$^{-1}$, respectively, i.e. they are much lower than the
rotational and collisional energies. Thus, assuming that the
electronic and nuclear spins play a spectator role during
electron-molecule collisions, hyperfine-resolved rate coefficients can
be computed using the simple infinite-order-sudden (IOS)
approximation. Within this approximation, which is similar in spirit
to the ANR approximation, the pure rotational rate coefficients obey
the following equation \citep{corey83}:
\begin{equation}
k^{IOS}_{N \to N'}(T)=[N']\sum_L \left(\begin{array}{ccc}
N' & N & L \\ 
0 & 0 & 0
\end{array}\right)^{2} k^{IOS}_{0 \to L}(T),
\label{iosrot}
\end{equation}
where $[N']$ represents $(2N'+1)$ and $\left( \quad \right)$ is a
Wigner ``3-j'' symbol. In practice, the rate coefficients $k_{N\to
  N'}(T)$ computed with \texttt{ROTIONS} do not strictly follow
Equ.~(\ref{iosrot}) due to the Coulomb-Born completion and the
threshold correction applied to the cross
sections. Equ.~(\ref{iosrot}) is however satisfied to within 25\%,
down to 10~K. Within the IOS approximation, the fine-structure rate
coefficients can be obtained as follows \citep{corey83,lique16}:
\begin{eqnarray}
\label{fs}
k^{IOS}_{Nj \to N'j'} (T) & = & [N N' j'] \sum_{L}  
\left(\begin{array}{ccc}
N' & N & L \\
0 & 0 & 0 
\end{array}
\right)^2 \nonumber \\
& & \times \left\{
\begin{array}{ccc}
N & N' & L \\
j' & j & S 
\end{array}
\right\}^2 \nonumber \\
& & \times k_{0\to L}(T), 
\end{eqnarray}
where $\left\{ \quad \right\}$ is a ``6-j'' Wigner symbol and $k_{0\to
  L}(T)$ are the rotational rate coefficients computed with
\texttt{ROTIONS}. Similarly, the hyperfine-resolved rate coefficients
can be obtained as \citep{daniel05,lique16}:
\begin{eqnarray}
\label{hfs}
k^{IOS}_{NjF \to N'j'F'} (T)  & = & [N N' j j' F'] 
\sum_{L} \left( 
\begin{array}{ccc}
N' & N & L \\
0 & 0 & 0 
\end{array}
\right)^2
\nonumber \\
& & \times \left\{
\begin{array}{ccc}
N & N' & L \\
j' & j & S 
\end{array}
\right\}^2 \nonumber \\
& & \times \left\{
\begin{array}{ccc}
j & j' & L \\
F' & F & I 
\end{array}
\right\}^2 k_{0\to L}(T).
\end{eqnarray}
In practice, however, the hyperfine rate coefficients for transitions
with $N\ne N'$ were computed as \citep{neufeld94,faure12}:
\begin{equation}
  k^{INF}_{NjF \to N'j'F'}(T) = \frac{k^{IOS}_{NjF \to
      N'j'F'}(T)}{k^{IOS}_{N\to N'}(T)}k_{N\to N'}(T).
  \label{scal}
\end{equation}
This scaling procedure guarantees the following equality:
\begin{equation}
  \sum_{j'F'} k^{INF}_{NjF\to N'j'F'}(T) = k_{N\to N'}(T),
\end{equation}
thus ensuring that the summmed hyperfine rate coefficients are
identical to the ANR/Coulomb-Born pure rotational rate
coefficients. In addition, in order to improve the results at low
temperatures, the fundamental excitation rate coefficients $k_{0\to
  L}(T)$ were replaced by the de-excitation fundamental rate
coefficients using the detailed balance relation (within the IOS
approximation) $k_{0\to L}(T)=[L]k_{L\to 0}(T)$, as in \cite{faure12}.

\section{Results}

There are no previous studies on these systems against which we can
compare. We start by considering results for electron-impact
excitation of OH$^+$.

\subsection{OH$^+$}  

\begin{figure}
\begin{center}
	\includegraphics*[scale=0.35]{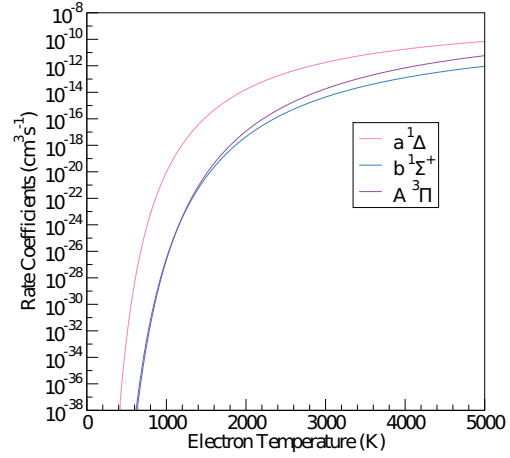}
\end{center}
	\caption{Rate coefficients for electronic excitation of
          OH$^+$.}
	\label{fig:OHPEER}
\end{figure}

Fig.~\ref{fig:OHPEER} shows the rate coefficients for the electronic
excitation of OH$^+$(X~$^3\Sigma^-$) after electron impact. This
figure shows that the excitation of OH$^+$(X~$^3\Sigma^-$) to
OH$^+$(a~$^1\Delta$) has a lower temperature threshold than the
subsequent transitions and has a greater magnitude over the
investigated temperature range. This is to be expected due to the
electron energy threshold of this transition, as shown in
Table~\ref{tab:OH+ExStat}. This figure also shows that while the rate
coefficients for excitation to OH$^+$(b~$^1\Sigma^+$) and
OH$^+$(A~$^3\Pi$) have a similar temperature threshold, the rate
coefficient for excitation to OH$^+$(A~$^3\Pi$)  dominates at
higher temperatures and in fact is converging towards the rate
coefficient for excitation to OH$^+$(a~$^1\Delta$).  This is 
a consequence of the fact that  the
OH$^+$(X~$^3\Sigma^-$) to OH$^+$(A~$^3\Pi$) transition is dipole
allowed so this excitation tends
to dominate at high impact energies.. State-to-state Einstein
coefficients for the $^3\Sigma^--^3\Pi$ band can be found in
\cite{gomez14}.

\begin{figure}
\begin{center}
	\includegraphics*[scale=0.35]{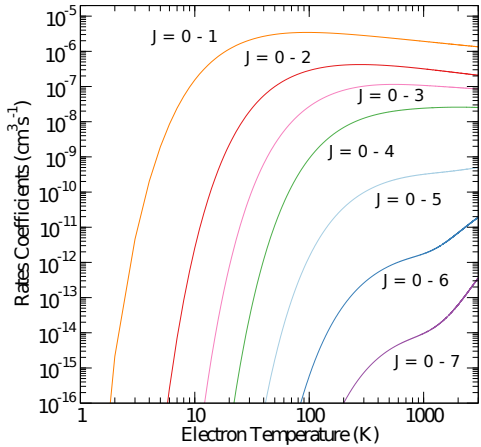}
\end{center}
	\caption{Rate coefficients for rotational excitation of OH$^+$
          from the ground state ($N$=0) to the lowest seven excited
          states.}
	\label{fig:OHP}
\end{figure}

Fig.~\ref{fig:OHP} presents rate coefficients for electron-impact
rotational excitation of OH$^+$ from its rotational ground state. The
processes are dominated by the $\Delta N = 1$ transition due to the
long-range effect of the dipole moment discussed above. As $\Delta N$
increases the temperature threshold of the process increases and the
magnitude of the rate coefficients decreases.

Table~\ref{tab:OH+HFS} presents rate coefficients for electron-impact
hyperfine de-excitation of OH$^+$ from the initial levels $(N, J,
F)=(1, 2, 5/2)$ and $(1, 2, 3/2)$. These two levels are the upper
states of the observed transition of OH$^+$ at 972~GHz that will be
discussed in the next section. It can be noticed that transitions with
$\Delta F=\Delta j=\Delta N=\pm 1$ are collisionally favored, as
observed previously for other $^3\Sigma^-$ targets colliding with
neutrals \citep[see][and references therein]{lique16}. We note that
radiatively the selection rules $\Delta F=0, \pm 1$ holds strictly and
transitions with $\Delta F=\Delta j=\Delta N$ are the strongest
ones. We also observe that de-excitation rate coefficients decrease
significantly with temperature, typically by a factor of 10 between 10
and 1000~K.

\begin{table}
\begin{center}
\caption[Hyperfine de-excitation rate coefficients for
  OH$^+$.]{Hyperfine de-excitation rate coefficients in cm$^3$s$^{-1}$
  for OH$^+$ in initial levels $(N, J, F)=(1, 2, 5/2)$ and $(1, 2,
  3/2)$. Powers of ten are given in parentheses.}\label{tab:OH+HFS}

\begin{tabular}{lllllllll}
  \hline
  \hline
N & j & F & N' & j' & F' & 10~K & 100~K & 1000~K \\
\hline
1 & 2 & 5/2 & 0 & 1 & 3/2 &  5.38(-6) & 1.72(-6) & 6.39(-7) \\
1 & 2 & 5/2 & 0 & 1 & 1/2 &  0.0      & 0.0      & 0.0 \\
1 & 2 & 5/2 & 1 & 0 & 1/2 &  3.81(-7) & 1.16(-7) & 3.80(-8) \\
1 & 2 & 3/2 & 0 & 1 & 3/2 &  8.97(-7) & 2.87(-7) & 1.06(-7)  \\
1 & 2 & 3/2 & 0 & 1 & 1/2 &  4.48(-6) & 1.43(-6) & 5.32(-7)  \\
1 & 2 & 3/2 & 1 & 0 & 1/2 &  3.81(-7) & 1.16(-7) & 3.80(-8)  \\
1 & 2 & 3/2 & 1 & 2 & 5/2 &  2.00(-7) & 6.09(-8) & 1.99(-8)  \\
\hline
\hline
\end{tabular}
\end{center}
\end{table}

\subsection{SH$^+$}

\begin{figure}
\begin{center}
	\includegraphics*[scale=0.35]{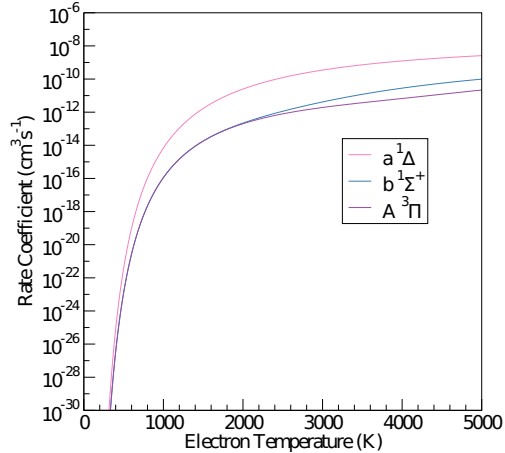}
\end{center}
	\caption{Rate coefficients for electronic excitation of
          SH$^+$.}
	\label{fig:SHPEER}
\end{figure}

Fig.~\ref{fig:SHPEER} shows the rate coefficients for the electronic
excitation of SH$^+$(X~$^3\Sigma^-$) after electron impact. This
figure shows that the temperature thresholds of the three transitions
considered in this work are fairly similar. 
The
rate coefficient for the transition to SH$^+$(a~$^1\Delta$) 
dominates from relatively low temperatures whereas the rate
coefficients for transitions to SH$^+$(b~$^1\Sigma^+$) and
SH$^+$(A~$^3\Pi$) remain very similar up to around 2000~K. At higher
temperatures, the rate coefficient for the transition to
SH$^+$(b~$^1\Sigma^+$) exceeds that for the transition to
SH$^+$(A~$^3\Pi$). This latter does however tend to converge towards
the former as the temperature increases still further.

\begin{figure}
\begin{center}
	\includegraphics*[scale=0.35]{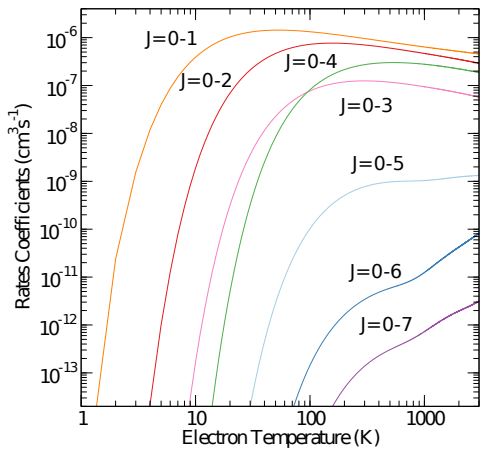}
\end{center}
	\caption{Rate coefficients for rotational excitation of
          SH$^+$ from the ground state ($N$=0) to the lowest seven
          excited states.}
	\label{fig:SHP}
\end{figure}

Fig.~\ref{fig:SHP} presents rate coeffcients for electron-impact
rotational excitation of SH$^+$ from its rotational ground state.  The
processes are again dominated by the $\Delta N = 1$ transition,
particularly at low temperatures. As $\Delta N$ increases the
temperature threshold of the process increases and the magnitude of
the rate coefficient decreases with the exception of the rate
coefficient for the $\Delta N = 4$ transition which comes to exceed
that of the $\Delta N = 3$ transition above $\sim$90~K.

Table~\ref{tab:SH+HFS} presents rate coefficients for electron-impact
hyperfine de-excitation of SH$^+$ from the initial levels $(N, J,
F)=(1, 2, 5/2)$ and $(1, 2, 3/2)$. These two levels are the upper
states of the transition of SH$^+$ at 526~GHz first detected with {\it
  Herschel} \citep{benz2010}. Again we can notice that transitions
with $\Delta F=\Delta j=\Delta N=\pm 1$ are favoured and that
de-excitation rate coefficients decrease by a factor of $\sim$10
between 10 and 1000~K.

\begin{table}
\begin{center}
\caption[Hyperfine de-excitation rate coefficients for
  SH$^+$.]{Hyperfine de-excitation rate coefficients in cm$^3$s$^{-1}$
  for SH$^+$ in initial levels $(N, J, F)=(1, 2, 5/2)$ and $(1, 2,
  3/2)$. Powers of ten are given in parentheses.}
\label{tab:SH+HFS}
\begin{tabular}{lllllllll}
  \hline
  \hline
N & j & F & N' & j' & F' & 10~K & 100~K & 1000~K \\
\hline
1 & 2 & 5/2 & 0 & 1 & 3/2 &  1.80(-6) & 5.75(-7) & 2.18(-7) \\
1 & 2 & 5/2 & 0 & 1 & 1/2 &  0.0      & 0.0      & 0.0 \\
1 & 2 & 5/2 & 1 & 0 & 1/2 &  4.04(-7) & 1.27(-7) & 4.04(-8) \\
1 & 2 & 3/2 & 0 & 1 & 3/2 &  3.00(-7) & 9.58(-8) & 3.63(-8)  \\
1 & 2 & 3/2 & 0 & 1 & 1/2 &  1.50(-6) & 4.79(-7) & 1.82(-7)  \\
1 & 2 & 3/2 & 1 & 0 & 1/2 &  4.04(-7) & 1.27(-7) & 4.04(-8)  \\
1 & 2 & 3/2 & 1 & 2 & 5/2 &  2.12(-7) & 6.68(-8) & 2.12(-8)  \\
\hline
\hline
\end{tabular}
\end{center}
\end{table}

The supplementary data associated with this paper include:
\begin{itemize}
\item Electronic excitation cross sections and rate coefficients for
  $^{16}$OH$^+$ and $^{32}$SH$^+$. Data include all electronic states
  with thresholds below 5~eV.
\item Rotation excitation cross sections and rate coefficients for the
  three isotopes of OH$^+$ and the four isotopes of SH$^+$. Rotational
  excitation datasets are published for transitions with starting
  values of $N = 0$ to $N = 11$.
\item Hyperfine de-excitation rate coefficients for $^{16}$OH$^+$,
  $^{18}$OH$^+$ and $^{32}$SH$^+$. Hyperfine de-excitation datasets
  are published for transitions with starting values of $(N, j, F) =
  (0, 1, 3/2)$ to $(11, 11, 21/2)$.
\end{itemize}
These data will also be placed in the BASECOL database \citep{jt547s}.

Hyperfine data for $^{16}$OH$^+$, $^{18}$OH$^+$ and
$^{32}$SH$^+$ have been also combined with the spectroscopic data from the
Cologne Database for Molecular Spectroscopy \citep{cdms} in order to
provide a full and consistent dataset adapted to radiative transfer
studies (see below). Hyperfine data for the other isotopologues are
not provided due to the lack of spectroscopic data (the recent entry
$^{34}$SH$^+$ at CDMS is currently limited to nine hyperfine
transitions within $N=1-0$).

\section{OH$^+$ excitation in the Orion bar}

The first detection of OH$^+$ in emission in a Galactic source was
reported by \cite{tak2013} using the {\it Herschel Space
  Observatory}. These authors presented line profiles and maps of
OH$^+$ line emission toward the Orion Bar PDR. The Orion Bar PDR is
the archetypal edge-on molecular cloud surface illuminated by
far-ultraviolet radiation from nearby massive stars. The analysis of
the chemistry and excitation of OH$^+$ by \cite{tak2013} suggests an
origin of the emission at visual extinctions $A_V\sim 0.1-1$ where
most of the electrons are provided by the ionized carbon atoms and
hydrogen is predominantly in atomic form. This is also the region
where CH$^+$ and SH$^+$ emissions originate \citep{nagy13}. In such an
environment, the dominant formation pathway for OH$^+$ is O$^+$ +
H$_2$ and the main destruction route is OH$^+$+H$_2$
\citep{tak2013}. The reaction of OH$^+$ with H is
endothermic. Chemical pumping may thus play a role in the excitation
of OH$^+$ only if the molecular fraction $f({\rm H_2})=2N({\rm
  H_2})/(2N({\rm H_2})+N({\rm H}))$ is large enough. Given that
$f({\rm H_2})$ is expected to be low ($<10\%$) in the PDR layers where
OH$^+$ ions form, the impact of chemical pumping should be small, as
found by \cite{gomez14}. This is in contrast with CH$^+$ which reacts
rapidly with H to form C$^+$ + H$_2$ \citep{jt688}.

We have thus assumed that the excitation of OH$^+$ is entirely driven
by inelastic collisions with electrons and hydrogen atoms. The
hyperfine collisional data presented above for OH$^+$ + $e^-$ and
those of \cite{lique16} for OH$^+$ + H were combined with
spectroscopic data from CDMS and implemented in a non-LTE radiative
transfer model. We have employed the public version of the
\texttt{RADEX}
code\footnote{http://home.strw.leidenuniv.nl/$\sim$moldata/radex.html}
which uses the escape probability formulation assuming an isothermal
and homogeneous medium. The cosmic microwave background (CMB) is the
only background radiation field with a temperature of
2.73~K. Radiative pumping by local dust and starlight is neglected in
order to focus on collisional excitation effects. We assume that
OH$^+$ probes a homogeneous region corresponding to the ``hot gas at
average density'' described by \cite{nagy17} for the Orion Bar: the
atomic hydrogen density is taken as $n({\rm H})=2\times
10^5$~cm$^{-2}$ and the kinetic temperature as $T_k=500$~K, that is a
thermal pressure of $10^8$~K.cm$^{-3}$ which is typical of dense
PDR. We adopted a typical electron fraction $x(e)=n(e^-)/n({\rm
  H})=10^{-4}$, as expected if carbon is fully ionized. The line width
was fixed at 4~km.s$^{-1}$, as observed by \cite{tak13}. Assuming a
unit filling factor, the OH$^+$ column density is the single free
parameter adjusted to best reproduce the integrated intensities
measured by \cite{tak13}. We have employed the three transitions
observed by these authors at 909.159, 971.804 and 1033.119~GHz,
corresponding to the transitions $(N, j, F)=(1, 0, 1/2)\to (0, 1,
3/2)$, $(1, 2, 5/2)\to (0, 1, 3/2)$ and $(1, 1, 3/2)\to (0, 1, 3/2)$,
respectively, which are the strongest hyperfine components in each
fine-structure line. It must be noted that the transition $(N, j,
F)=(1, 2, 5/2)\to (0, 1, 3/2)$ is actually blended with the transition
$(1, 2, 3/2)\to (0, 1, 1/2)$ at 971.805~GHz. Since \texttt{RADEX} does
not treat the overlap of lines, it was necessary to extract the
excitation temperature and line center opacity of the blended
transitions. Assuming Gaussian shapes, the opacities were summed to
simulate a composite line whose intensity was integrated over velocity
range from -10 to +10~km.s$^{-1}$. Overlap effects should be properly
included in the radiative transfer treatment but given the low opacity
of the lines ($\tau <2$) their impact is expected to be moderate here.

\begin{figure}
\begin{center}
\rotatebox{90}{\includegraphics*[width=7.5cm,angle=180]{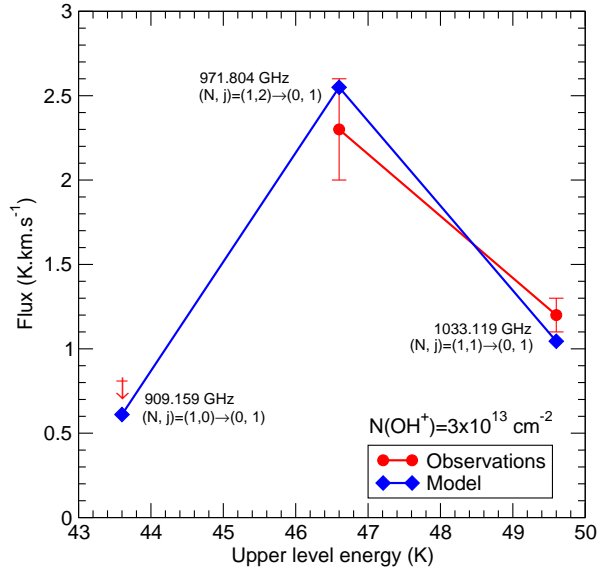}}
\end{center}
	\caption{OH$^+$ line flux of the strongest hyperfine
          components of the transition $(N, j)=(1, j)\to (0, 1)$ as
          functions of the upper level energy, as predicted by our
          non-LTE calculations for the physical conditions used for
          the Orion Bar. The OH$^+$ column density was adjusted to
          best reproduce the observations of van der Tak et
          al. (2013). See text for details.}
	\label{fig:orion}
\end{figure}

Very good agreement is observed in Fig.~\ref{fig:orion} between our
model and the observations for a OH$^+$ column density of $3\times
10^{13}$~cm$^{-2}$. Indeed, the calculations agree, essentially within
error bars, with {\it Herschel} data at 971.804 and 1033.119~GHz. They
are also consistent with the upper limit at 909.159~GHz. Our column
density is a factor of $\sim 3$ lower than the value derived by
\cite{tak13}. These authors have employed similar physical conditions
but different collisional data and they included chemical terms, which
explains the difference. On the other hand, we note that our result is
in good agreement with the column density derived by \cite{tak13}
using the abundance predicted by the Meudon PDR code ($1.6\times
10^{13}$~cm$^{-2}$). Finally, the contribution of electron collisions
was found to be moderate, of the order of 10-20\%, at an electron
fraction $x_e=10^{-4}$. The excitation of OH$^+$ in the Orion Bar is
therefore dominated by hydrogen collisions. The impact of
electron-impact excitation would be much larger in environments with
high ionisation fractions such as supernova remnants
\citep{barlow13,jt617} or planetary nebulae \citep{aleman14}.

\section{Conclusions}

Electronic and rotational excitation cross sections and rate
coefficients have been produced and made available for a range of
rotational transitions of the open-shell hydrides OH$^+$ and SH$^+$
and their isotopologues. The electronic structure calculations were
validated where possible against published data. The calculated
excitation thresholds, calculated dipole transition moments and
rotational constants of both hydrides were validated against measured
values or values recommended by the CDMS \citep{cdms} and these
comparisons are very good.

The \textbf{R}-matrix method was used to calculate \textbf{T}-matrices
from which electronically and rotationally inelastic cross sections
were calculated. No published data were available to validate these
inelastic cross sections but the reliability of the ANR/Coulomb-Born
approach was previously confirmed both experimentally and
theoretically. Rate coefficients were calculated by integration of the
cross sections using Maxwell-Boltzman distribution of electron
velocities. Hyperfine de-excitation rate coefficients were deduced
from the rotational data using the IOS approximation. As with the
closed shell hydrides \citep{jt617}, the rotational excitation rate
coefficients of the $\Delta N = 1$ transitions were found to be
strongly influenced by the long-range effect of the dipole moment and
have the largest magnitudes. This result was found to translate in the
hyperfine propensity rule $\Delta F=\Delta j=\Delta N=\pm 1$.

The electron-impact excitation data were combined with the results of
\cite{lique16} for OH$^+$+H collisions in order to model the
rotational/hyperfine excitation of OH$^+$ in the Orion Bar PDR. Very
good agreement with the observations of \cite{tak13} was obtained for
a OH$^+$ column density of $3\times 10^{13}$~cm$^{-2}$, which is
similar to the prediction of the Meudon PDR model. We recommend using
the present data in any model of OH$^+$ excitation in regions where
the electron fraction is larger than 10$^{-4}$. 

Finally, electron collisions can seed processes besides rotational
excitation and electronic excitation. For molecular ions both dissociative
recombination (DR) and vibrational excitation can be astrophysically important  processes.
The mechanisms for these differ somewhat from that considered above as their cross
sections are  dominated by the contribution of resonances. They thus require
rather more extensive theoretical procedures, see for example \citet{jt591}. We note that
electron-impact vibrational excitation and DR rate coefficients have very recently been
computed by \citet{stroe18}.

\section*{Acknowledgements}

We thank  Yohann Scribano for help at the start of this project and
Fran\c{c}ois Lique for helpful discussions.
This work has been supported by an STFC CASE studentship, grant number
ST/K004069, for JRH, the Agence Nationale de la Recherche
(ANR-HYDRIDES), contract ANR-12-BS05-0011-01 and by the CNRS national
program ``Physico-Chimie du Milieu Interstellaire".

\bibliographystyle{mn2e}

\end{document}